\documentclass{aa}
\usepackage{txfonts}
\usepackage{ifthen}
\usepackage{graphicx}
\usepackage{supertabular}

\newcommand{\Msun}{M$_{\odot}$}

\newcommand{\Lsun}{L$_{\odot}$}

\newcommand{\Day}{$^{d}$}

\newcommand{\Mag}{$^{m}$}

\newcommand{\mic}{$\mu$m}

\newcommand{\Mdot}{$\dot{\rm{M}}$}

\newcommand{\Mms}{M$_{\rm{MS}}$}

\newcommand{\Myr}{M$_{\odot}$year$^{-1}$}

\begin{document}

\title {An infrared study of galactic OH/IR stars. I.\\ An
 optical/near-IR atlas of the Arecibo sample\thanks{Based on
 observations collected at the German-Spanish Astronomical Centre,
 Calar Alto, operated jointly by Max-Planck-Institut f\"uer Astronomie
 and Instituto de Astrof\'\i sica de Andaluc\'\i a (CSIC).}$^,$
 \thanks{Tables A1 and A2 are only available in electronic form at the
 CDS via anonymous ftp to cdsarc.u-strasbg.fr (130.79.128.5) or via
 http://cdsweb.u-strasbg.fr/cgi-bin/qcat?J/A+A/}$^,$\thanks{The
 complete atlas can be accessed electronically at:
 http://www.edpsciences.org/aa}}

\author {F.M. Jim\'enez-Esteban \inst{1,2} 
  \and L. Agudo-M\'erida  \inst{1} 
  \and  D. Engels \inst{2}
  \and P. Garc\'{\i}a-Lario \inst{3}
}

\institute{VILSPA Satellite Tracking Station, Apartado de Correos 50727, E-28080 Madrid, Spain
  \and Hamburger Sternwarte, Gojenbergsweg 112, D-21029 Hamburg, Germany 
  \and ISO Data Centre, Science Operations and Data Systems Division, Research
 and Scientific Support Department of ESA, Villafranca del Castillo,  
 Apartado de Correos 50727, E-28080 Madrid, Spain
}

\offprints{F.M. Jim\'enez-Esteban\\
\email{Francisco.Jimenez-Esteban@hs.uni-hamburg.de}}

\date{Received 22 July 2004 / Accepted 11 October 2004}	

\titlerunning{An optical/near-IR atlas of OH/IR stars}
\authorrunning{Jim\'enez-Esteban et al. }


\abstract{
In this paper we present optical and near-infrared finding charts,
accurate astrometry ($\approx$\,1\arcsec) and single-epoch
near-infrared photometry for 371 IRAS sources, 96\%
of those included in the so-called `Arecibo sample of OH/IR stars'
(Eder et al. \cite{Eder88}; Lewis et al. \cite{Lewis90a}; Chengalur et
al. \cite{Chengalur93}). The main photometric properties of the stars
in the sample are presented and discussed as well as the problems
found during the process of identification of the
optical/near-infrared counterparts. In addition, we also identify
suitable reference stars in each field to be used for differential
photometry purposes in the future.

We find that 39\% of the sources (144 in number) have no optical
counterpart, 8 of them being invisible even at near infrared
wavelengths. The relative distribution of sources with and without
optical counterpart in the IRAS two-colour diagram and their
characteristic near infrared colours are interpreted as the
consequence of the increasing thickness of their circumstellar shells.
Among the objects not detected at near infrared wavelengths four
non-variable sources are proposed to be heavily obscured post-AGB
stars which have just very recently left the AGB.
Eight additional objects with unusually bright and/or blue
near-infrared colours are identified as candidate post-AGB stars
and/or proto-planetary nebulae.

}

\maketitle

\section {Introduction}


Stars with masses between 0.8 and 8 M$_{\odot}$ pass along the
Asymptotic Giant Branch (AGB) towards the end of their
evolution. Approaching the tip of the AGB they start to pulsate and
appear as large-amplitude ($\ga$\,0.4\Mag\, at K-band) long-period
($>$\,80\,days) variables. The pulsation is accompanied by heavy mass
loss which forms a circumstellar envelope of gas and dust. If the mass
loss rate surpasses \Mdot\,$\ge$\,10$^{-6}$\,\Myr\, the dust shell
eventually becomes opaque to visible light (Habing \cite{Habing96}).

Stellar evolution models predict that stars close to the tip of the
AGB experience thermal pulses (the so-called `TP-AGB') leading to
variations in luminosity and mass loss rates on timescales of several
ten thousand years. In addition, they may change their photospheric
chemistry from oxygen- to carbon-rich. The determination of the
current evolutionary state of a particular AGB star is difficult
because of the strong variability and the increasing obscuration which
limits our observational capabilities. Moreover, their pulsation on
timescales of 1\,--\,5\,years usually requires long observational
monitoring programs to determine representative parameters. Hence,
well defined AGB star samples, observed over an adequate long period
of time, are required to test current evolutionary models of this
short-lived phase of stellar evolution.

Available samples of AGB stars are in many cases biased. Depending on
whether the selection is made at optical/near-infrared wavelengths or
in the far-infrared the samples under analysis contain a larger number
of low-mass or high-mass AGB stars, respectively. Low-mass progenitors
are expected to experience only moderate mass loss during their AGB
lifetimes and, thus they will be detected preferentially in the
optical and near-infrared. An example is the `dust-enshrouded AGB
sample' of Jura \& Kleinmann (\cite{Jura89}), which contains stars
with typical main-sequence masses \Mms\,$\approx$\,1.3\,\Msun\,
(Olivier et al. \cite{Olivier01}). In contrast, AGB stars descending
from progenitors with larger main-sequence masses
(\Mms\,$\ga$\,2\,M$_{\odot}$) probably spend a larger fraction of
their TP-AGB life in an obscured state. This is a consequence of the
higher mass loss rates, which favours their detection in the
far-infrared. Less biased AGB samples covering the complete TP-AGB and
all kinds of chemistries should therefore include these far-infrared
sources, something which is possible only since the
infrared survey by the IRAS satellite was completed (Beichman et
al. \cite{Beichman88}).

Samples of oxygen-rich AGB stars (`OH/IR stars') can be constructed
using surveys of OH maser emission. In such surveys AGB stars are
easily picked up because of their conspicuous double-peaked maser
profile. However, these samples are also incomplete, as only part of
the oxygen-rich AGB stars exhibit OH maser emission (Lewis \& Engels
\cite{Lewis95}). Monitoring of early samples (e.g. Baud et
al. \cite{Baud81}) revealed either very long periods or rather weak
variations (Engels et al. \cite{Engels83}; Herman \& Habing
\cite{Herman85}). These OH/IR stars were considered as massive
AGB stars (\Mms\,=\,3\,--\,8\,\Msun\, on the main sequence), with the
long-period variables shortly before and the non-variables probably
shortly after the departure from the AGB. From the analysis of ISO
data of galactic AGB and post-AGB stars, Garc\'\i a-Lario \&
Perea-Calder\'on (\cite{Garcia-Lario04}) have recently postulated
the existence of three different chemical branches of AGB stars,
depending on their progenitor mass: (i) low-mass oxygen-rich stars;
(ii) intermediate-mass carbon-rich stars; and (iii) high-mass O-rich
stars, where `hot bottom burning' is active at the base of the
convective envelope, preventing the formation of carbon (instead, the
production of nitrogen is favoured). It is however not yet settled if
low-mass oxygen-rich AGB stars evolve into the large mass loss regime
observed in high-mass oxygen-rich AGB stars, and if so, how long they
stay in this regime. It is therefore not clear what fraction of
low-mass stars is present in OH/IR star samples. Actually, the
relation between AGB star samples selected in different wavelength
ranges remains to be explored.

\begin{figure*}
\begin{center}
   \includegraphics[width=13cm]{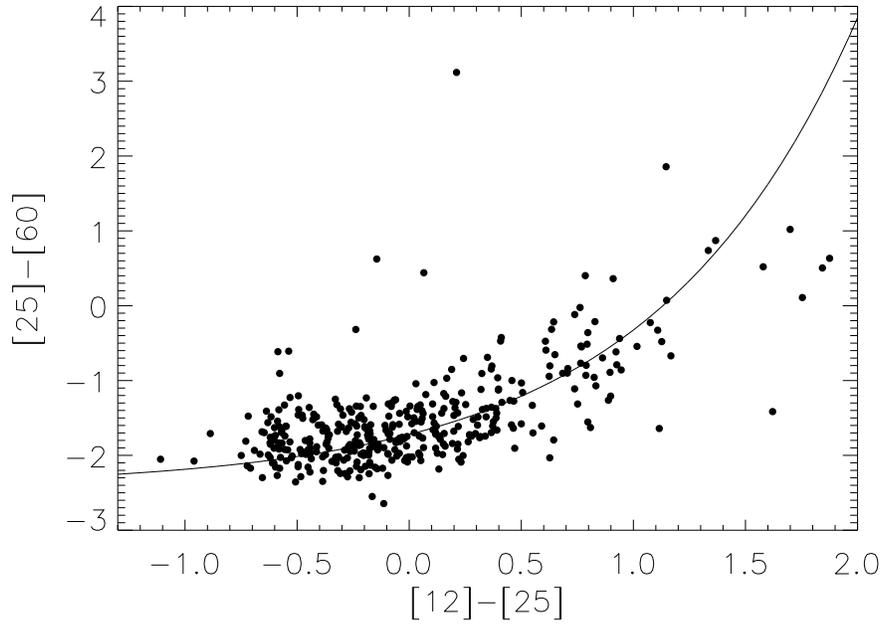}
    \caption[IRAS colour-colour diagram for the `Arecibo
             sample']{\label{fig_arec:IRAS_colour}\,\, The position in
             the IRAS two-colour diagram of the stars in the `Arecibo
             sample', where the IRAS colours are defined as:
             [12]$-$[25]\,=\,$-$2.5\,log$\frac{F_{\nu}(12)}{F_{\nu}(25)}$
             and
             [25]$-$[60]\,=\,$-$2.5\,log$\frac{F_{\nu}(25)}{F_{\nu}(60)}$. The
             continuous line is what we call the `O-rich AGB sequence'
             (see text).}
\end{center}
\end{figure*}


A well-defined sample of far-infrared-selected oxygen-rich AGB stars
is provided by the Arecibo survey of OH/IR stars (Eder et
al. \cite{Eder88}; Lewis et al. \cite{Lewis90a}; Chengalur
et al. \cite{Chengalur93}). It consists of $\approx$\,390 IRAS
sources, which were detected in the 1612\,MHz OH maser line with the
Arecibo radio telescope. The sample was obtained from a complete
survey of IRAS sources with flux densities $\ge$\,2\,Jy at 25\,\mic,
with declination 0\degr\,$<$\,$\delta$\,$<$\,37\degr\, and appropriate
colours of AGB stars (Olnon et al. \cite{Olnon84}). The OH maser
detection qualifies the IRAS source as an oxygen-rich AGB star. The
positions of the sources in the IRAS two-colour diagram are shown in
Fig. \ref{fig_arec:IRAS_colour}, together with a solid line which
corresponds to the sequence of colours predicted for oxygen-rich AGB
stars with increasing mass loss (Bedijn \cite{Bedijn87}) that
we have named the `O-rich AGB sequence'. The broad range of
colours suggests that this sample bridges the gap between the
optically/near-infrared selected AGB samples and the OH/IR stars
discovered by OH maser surveys.

The Arecibo sample is probably composed of objects with a variety of
variability properties, although the Mira-type large amplitude
variables are thought to prevail. Unfortunately, variability
information only exists in the literature for around 50 sources in the
sample, concentrated at the blue end of the IRAS two-colour
diagram. 
Most of them are well known Mira variables
(145\Day\,$<$\,P\,$<$\,660\Day) listed as such in the General Catalogue of
Variable Stars (Kholopov et al. \cite{Kholopov98}) on the basis of
optical observations. Ten of them have also been monitored in the
near-infrared (Whitelock et al. \cite{Whitelock94}; Olivier et
al. \cite{Olivier01}), displaying amplitudes of
0.5\Mag\,$<$\,$\Delta$K\,$<$\,1.7\Mag. 
At least one of these blue sources, IRAS\,02404+2150, has been
reported to show a low amplitude semiregular variability (Whitelock et
al. \cite{Whitelock95}). At the red end a handful of sources has
previously been referred to in the literature either as variable or
non-variable OH/IR stars, but usually a detailed knowledge of their
light curves is still lacking. 
In addition, a few peculiar sources are also part of the sample, such
as the very young planetary nebula M 1$-$92 (IRAS\,19343+2926).

In order to study the variability properties of the `Arecibo sample'
(periods, amplitudes, and colour variations) we started in 1999 a
near-infrared monitoring program that we are still carrying on. At the
end of 2003 we had already completed 11 observing runs with more
than 110 nights of observation. A subsample is also being monitored
optically. The first results of this multi-wavelength long-term
monitoring program are reported in this paper. Our ultimate goal is
to study in particular the oxygen-rich AGB stars with
\Mms\,$\ga$\,2\,\Msun\, which are probably rare among AGB stars
discovered optically and in the near-infrared, but are common in
samples discovered by blind OH maser surveys.

In this paper, which is the first of a series of three, we will focus
on the determination of the optical/near-infrared counterparts of all
the OH/IR stars in the `Arecibo sample' and on the identification of
suitable reference stars in each field to be used for differential
photometry purposes in the future. The result is an atlas of finding
charts combining optical images from the {\it Digitized Sky Survey}
(Djorgovski et al. \cite{Djorgovski01}) and our own near-infrared
images, obtained in the J, H, and K bands.

Sect. 2 describes the observations performed and the acquisition and
data analysis techniques applied. In Sect. 3 we explain how we
identified the optical/near-infrared counterparts and how we
determined new accurate coordinates. This is followed by a brief
description of the atlas. In Sect. 4 we present the near-infrared
photometry, while the results obtained are discussed in Sect.
5. Finally, the main conclusions are presented in Sect. 6.

\section {\label{Observations} Observations and data reduction}


Our `Arecibo sample' of OH/IR stars contains 385 objects, and
comprises most of the original sources listed in Eder et
al. (\cite{Eder88}), Lewis et al. (\cite{Lewis90a}), and
Chengalur et al. (\cite{Chengalur93}). The only objects not
considered here were sidelobe responses or serendipitous OH maser
discoveries, as described in Engels \& Lewis (\cite{Engels96}),
together with three objects (IRAS\,18534+0215, IRAS\,19175+1344, and
IRAS\,19226+1401) that were left out because they were re-classified
by Lewis (\cite{Lewis97}) as molecular cloud sources. Finally
IRAS\,19200+2101 was also removed, as it turned out to be an erroneous
entry in the IRAS PSC (see Lewis et al. \cite{Lewis04}).

\subsection{Observations}
The infrared observations under discussion here were performed during
20 nights in July 1999 and July 2000 at Calar Alto Observatory
(Almer\'{\i}a, Spain) using the 1.23\,m telescope equipped with the
infrared camera MAGIC (Herbst et al. \cite{Herbst93}). All sources in
the sample, except 13 objects not accessible at this time of the year
(with 4$^h$\,$<$\,$\alpha$\,$<$\,12$^h$) and the very bright, detector
saturating, Mira variable R\,Aql (IRAS\,19039+0809), were observed in
the J (1.25\,$\mu$m), H (1.65\,$\mu$m) and K (2.20\,$\mu$m) near
infrared bands. The sources observed (N\,=\,371) are listed in Table
A2 of the Appendix. All except one (IRAS\,18549+0208) were observed at
least once under photometric weather conditions. The MAGIC camera uses
a 256\,$\times$\,256 pixels NICMOS3 array that provides an approximate
field of 5\arcmin\,$\times$\,5\arcmin. Typical on-source integration
times were 1 minute for the J and H bands, and 30 seconds for the K
band, resulting in typical upper detection limits of around 15\Mag,
14\Mag\, and 13\Mag, respectively. To avoid saturation when observing
the brightest sources a small aperture configuration was
used. Standard stars from Elias et al.  (\cite{Elias82}) were observed
during each night for photometric calibration. They were followed over
a range of airmasses to determine the atmospheric extinction
corrections.

The main difficulty to observing in the infrared is the high sky
background level and its fast temporal and spatial variation. For the
accurate determination of the background we took for each object and
every filter 5 images with the object centered at different positions
in the image (dithering or `moving sky' technique). Thus, each
measurement consisted of a set of 5 individual images. Taking the
median of the 5 images we obtained a temporally and spatially
well determined sky image which was subtracted from the object
image. Flat-field images were also taken every night for each
individual filter in each of both aperture configurations to correct
for pixel-to-pixel sensitivity variations during the data reduction
process.

\subsection{Data reduction}
During each of the two observing epochs we obtained of the order of
ten thousand images covering the full sample twice. Within the
long-term monitoring project we expect to take $\approx$150\,000
images, which makes a high level of automation in data reduction
desirable. Thus, a semi-automated reduction procedure was developed
combining a self-written IDL (Interactive Data Language; Research
Systems Inc.) routine with pre-existing routines within the
IRAF\footnote{IRAF is distributed by the National Optical Astronomy
Observatories, which are operated by the Association of Universities
for Research in Astronomy, Inc., under cooperative agreement with the
National Science Foundation.} software package. The IDL routine first
reads in all images taken of in a particular observing night from the
observing logfile provided by the MAGIC camera software. Then it
writes an IRAF macro, which combines several IRAF routines to perform
the data reduction which consisted of the following steps. First, a
detector mask was used in order to remove the bad pixels from the
images. Then these images were divided by the appropriate flat-field
images and subsequently the median sky images were subtracted. The
sky-subtracted images were then inspected for cosmic ray hits and
cleaned by interpolating between the neighbouring pixels.  Finally,
each set of 5 images obtained per source and filter were aligned and
combined to yield the final near-infrared atlas image.

\section {Identification of the optical/NIR counterparts}


In order to determine the optical/near-infrared counterparts of the
sources in our sample we first considered the best coordinates
available from the literature or from existing catalogues and then
searched for plausible counterparts at these locations both on the
optical images from the {\it Digitized Sky Survey} and on our
near-infrared images.

Improved coordinates with respect to those originally provided by the
IRAS Point Source Catalogue (accuracy typically between
10\arcsec\,--\,15\arcsec) were obtained in many cases by
cross-correlating our sample with the MSX6C Point Source Catalogue
(Egan et al. \cite{Egan03}), which provides coordinates with an
accuracy of $\approx$\,2\arcsec. Although the MSX survey was limited
to low Galactic latitudes ($\le$\,6$^{\circ}$ in absolute value), its
accuracy is in many cases essential for identifying the near-infrared
counterpart, in particular in crowded regions along the galactic plane
and for extremely red objects only marginally detectable in the K
band, as we will see later.

Out of the 269 objects (73\% of the observed sample) located within the
region of the sky surveyed by MSX, (essentially
$\mid$b$\mid$\,$<$\,6$^{\circ}$), we found that 249 objects (93\%) had
a mid-infrared counterpart in the MSX6C Point Source Catalogue (see
Table \ref{tab_arec:no-counterpart} and Table A1 in the Appendix). The
missing objects are all located at the edges of the MSX survey area,
at galactic latitudes in the range
4.9$^{\circ}$\,$<$\,$\mid$b$\mid$\,$<$\,6$^{\circ}$. Two sources
(IRAS\,19161+2343 and IRAS\,19206+2517), although not included in the
Point Source Catalogue, were identified directly on the MSX images,
from which we also derived the improved coordinates that were later
used in the identification process.

Additional improved coordinates were taken from Lewis et
al. (\cite{Lewis90b}), who determined positions of the OH masers
by radio interferometry at the VLA for 46 sources in our list. These
coordinates have errors $<$\,1\arcsec. Note that part of these objects
are also MSX detections.



Once the best coordinates available were determined for each source in
the sample, we searched for their near-infrared counterparts in our
images obtained at Calar Alto. For this we inspected a
30\arcsec\,$\times$\,30\arcsec\ box (or smaller, depending on the
accuracy of the available - IRAS, MSX or VLA - coordinates) centered
at the nominal Arecibo source position, and searched for a plausible
(i.e. redder than average) counterpart within the box. Usually only
one `red' source was found close to the expected position; in many
cases this was the brightest source in the near-infrared field and/or
showed extremely red colours.
 
In parallel, we searched also the corresponding position in the Second
Digitized Sky Survey (DSS2) using the red filter image, which covers
the spectral range 6\,000\,--\,7\,000\,\AA\, with a maximum efficiency
around 6\,700\,\AA. In many cases it was easy to confirm the previous
identification made in the near-infrared since no optical counterpart
(or a very faint one) was seen at the corresponding position.

However, in several cases additional information was needed to verify
the optical and near-infrared identification. For a few Arecibo
sources more than one candidate was found with appropriate
colours. The right counterpart was then uniquely identified by
searching for variability between two observing epochs. The same
verification method was used for near-infrared counterparts which
were detected only in the K band, and for which therefore colour
information was lacking.

In most cases (61\%),
a single, point source counterpart was found both in the optical and
in the near-infrared. For less than one third of the sample (28\%),
a bright near-infrared counterpart was found but so heavily obscured
in the optical that nothing was seen on the DSS2 image above the
detection limit of $\approx$\,20.8\Mag. 11\% of the sources
are so strongly obscured that in the near-infrared the counterpart was
not identified above the detection limit either (2\%), or only in the K-band
(9\%), being completely invisible at shorter wavelengths.

In all but 8 cases the identification strategy provided a plausible
counterpart to the Arecibo source, either in the optical and in the
near-infrared or only in the near-infrared.  The 8 sources without
counterpart are listed in Table \ref{tab_arec:no-counterpart}. All
have MSX and/or VLA coordinates, and their counterparts must have been
fainter than our detection limits at the two epochs of observation
considered.  Two of these objects, IRAS\,18498$-$0017 (OH\,32.8$-$0.3,
a well known long-period variable OH/IR star, Engels et
al. \cite{Engels83}) and IRAS\,19440+2251 (Lawrence et
al. \cite{Lawrence90}) were reported as detected sources in the past
but only at wavelengths beyond 3\,\mic. They do not appear to have a
counterpart in the 2MASS Point Source Catalogue (Cutri et
al. \cite{Cutri03}) either. Two of the sources in Table
\ref{tab_arec:no-counterpart}, IRAS\,18517+0037 and
IRAS\,19374+1626, were classified as variable OH/IR stars in the GLMP
catalogue (Garc\'\i a-Lario \cite{Garcia-Lario92}), based on the large
value of their IRAS variability index (quoted in Column 9). The same
classification applies also to IRAS\,18475+0353 and the above
mentioned IRAS\,18498$-$0017. The remaining three objects have a low
variability index. One of them, IRAS\,18596+0315 (OH\,37.1$-$0.8,
GLMP\,862) is a non-variable OH/IR star which has already left the AGB
(Engels \cite{Engels02}). If the non- or small amplitude variability
of these objects is confirmed, they are probably also non-variable
OH/IR stars that have left the AGB recently.



\begin{table*}

\caption[]{\label{tab_arec:no-counterpart}Arecibo OH/IR stars without
optical/near-infrared counterpart. Coordinates are from radio
interferometry (N=2) or from MSX.}

\begin{center}
\begin{tabular}[t]{cllrrrrccl}
\hline\hline\noalign{\smallskip}
      &              &                  & \multicolumn{2}{c}{MSX$-$IRAS}  & \multicolumn{2}{c}{VLA--IRAS}   &          &     &             \\
 IRAS & \multicolumn{2}{c}{Coordinates} & $\Delta\alpha$ & $\Delta\delta$ & $\Delta\alpha$ & $\Delta\delta$ & MSX6C\_G & VAR & Other names \\
      & \multicolumn{2}{c}{(J2000)}     & [\arcsec]      & [\arcsec]      & [\arcsec]      & [\arcsec]      &          &     &             \\
\hline\noalign{\smallskip}\noalign{\smallskip}
\object{18475+0353}   & 18 50 00.5  & +03 56 33   &     6 &    1 &   &   & 036.2795+02.1143   & 99 & \\
\object{18498$-$0017} & 18 52 22.2  & $-$00 14 11 &  $-$2 & $-$2 &   &   & 032.8276$-$00.3152 & 96 & OH~32.8$$-$$0.3 \\
\object{18501+0013}   & 18 52 40.02 & +00 16 46.9 &     4 &    0 & 5 & 0 & 033.3208$-$00.1459 & 15 & \\
\object{18517+0037}   & 18 54 20.8  & +00 41 05   &  $-$2 &    3 &   &   & 033.8728$-$00.3350 & 85 & GLMP 842\\
\object{18596+0315}   & 19 02 06.3  & +03 20 16   &  $-$2 &    1 &   &   & 037.1185$-$00.8473 & 17 & OH~37.1$$-$$08, GLMP 862 \\
\object{19006+0624}   & 19 03 03.4  & +06 28 54   &     1 &    2 &   &   & 040.0220+00.3818   & 16 & \\
\object{19374+1626}   & 19 39 39.2  & +16 33 41   & $-$17 & $-$3 &   &   & 053.1355$-$02.7609 & 60 & GLMP 939\\
\object{19440+2251}   & 19 46 09.27 & +22 59 24.0 &     6 &    0 & 7 & 0 & 059.4784$-$00.8969 & 16 &  \\
\noalign{\smallskip}\hline
\end{tabular} 
\end{center}
\end{table*}

\normalsize


Taking into account that IRAS\,18498$-$0017 (OH\,32.8$-$0.3) has a
period P\,$>$\,1\,500\Day\, (Engels et al. \cite{Engels83}) and
that the other stars in Table \ref{tab_arec:no-counterpart} are among
the most obscured objects in the sample, we conclude that the variable
sources are probably OH/IR stars with very long periods
($>$\,1\,000\Day) at the very end of the AGB, while the non-variable
ones must be heavily obscured post-AGB stars, already in the
transition to the planetary nebula stage.

\subsection{\label{arec_sect_atlas}The atlas of optical and near-infrared counterparts}


\begin{figure*}
\begin{center}
	\includegraphics[width=12.6cm]{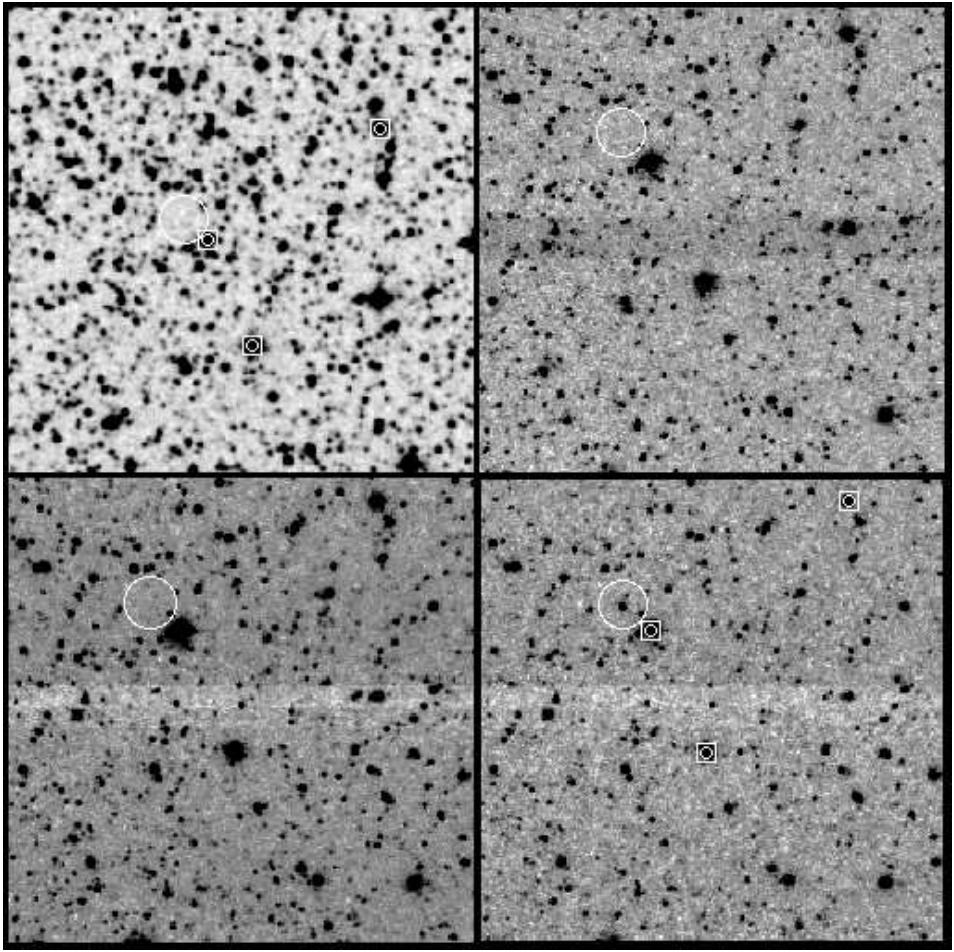} 

	\caption[Atlas images for
IRAS\,19323+1952]{\label{fig_arec:atlas_example} \,\, Atlas images for
IRAS\,19323+1952. The upper row gives the optical DSS2 R band and the
Calar Alto J band images, and the lower row the H- and K-band
images. The adopted counterpart is marked by a big circle, the
reference stars by small circles surrounded by squares.}

\end{center}  
\end{figure*}


The selected counterparts are displayed as an atlas of finding
charts. An example of the atlas images is given in Fig.
\ref{fig_arec:atlas_example}. For each Arecibo source in our sample a
chart was put together as a mosaic of 4 frames containing the optical
image taken from the DSS2 in the upper left panel, and the J, H and K
images from our observations in the upper right, lower left and lower
right panels, respectively. The size of the field shown in each
filter is 4.6\arcmin\,$\times$\,4.6\arcmin. For each source we marked
the position of the optical/near-infrared counterpart with a circle in
each of the available frames, as well as the position of the reference
stars used for the astrometric measurements (with small circles
surrounded by squares). These are the same stars which are used for
differential photometry purposes in our on-going long-term monitoring
program. In those cases where an optical/near-infrared counterpart was
not found the circle was drawn at the position where the source should
be located according to the best coordinates available (MSX or VLA).
The complete atlas can be accessed electronically at:
http://www.edpsciences.org/aa

\subsection{Derivation of improved coordinates}

Once the counterparts were identified, we used the set of reference
field stars (usually three) visible in both optical and near-infrared
images to derive new accurate coordinates of the target stars. Using
the relative distances between the Arecibo source and the reference
stars in the near-infrared we can determine the position of the
counterpart with an accuracy equivalent to less than one pixel on the
optical image. If a source was found at that precise position it was
considered to be the optical counterpart of the Arecibo source.

In a few cases (4\%)
the near-infrared counterpart was so bright that the short integration
time used to avoid saturation was insufficient to detect any reference
star in the field.  In these cases, a very bright object was also
visible at the nominal position in the DSS2 image and this was taken
as the optical counterpart. The plausibility of this identification is
supported by the blue IRAS colours associated with these extremely
bright near-infrared sources, which imply that they are not expected
to be heavily obscured in the optical. Because of the very bright
near-infrared counterpart of these stars
the variability monitoring is done in the optical to avoid saturation
problems during their luminosity maximum.

Taking advantage of the astrometric accuracy (1\arcsec/pixel) of the
DSS2 images we systematically improved the accuracy of the
astronomical coordinates of all sources in the sample for which either
an optical or a near-infrared counterpart was found. If the Arecibo
source had both an optical and a near-infrared counterpart, the
coordinates were taken directly from the DSS2 image by determining the
centroid of the point-like emission associated to the optical
counterpart. If only a near-infrared counterpart was found, the
position of the Arecibo source was determined from its relative
position to the reference stars in the near-infrared field, which were
always chosen to have point-like optical counterparts. In both cases,
the typical astrometric errors are estimated, from the pixel scale of
the DSS2 images, to be $\approx$\,1\arcsec\, both in RA and DEC.  The
complete list (363 objects)
of new coordinates is displayed in Table A1,
where we also list the relative distance from the
optical/near-infrared counterpart found to the original IRAS
coordinates. When available, the distance of the associated MSX and
VLA counterparts to the IRAS position is also provided for comparison.

If the Arecibo source had neither an optical nor a near-infrared
counterpart, then the coordinates assigned are the best ones available
for the source in the mid-infrared (MSX) or at radio wavelengths (for
the two with VLA measurements). These are listed in Table
\ref{tab_arec:no-counterpart}.

\section {Near infrared photometry}


Near infrared photometry was performed on each of the sources in the
sample for which a near infrared counterpart was found considering the
whole set of images available. The SEXtractor software package (Bertin
\& Arnouts, \cite{Bertin96}) was used in combination with IDL routines
to determine for each source the residual background noise $\sigma$ in
each of the five frames available per filter and to generate a
catalogue with the accurate position and the aperture photometry (in
integrated counts) of all the point sources detected in the field with
a signal at least a 3\,$\sigma$ above the background. The counts
corresponding to the source identified as the Arecibo counterpart were
then extracted from these files and averaged, and a standard error
$\sigma^\prime$ was calculated. Individual measurements exceeding the
average value by more than 3\,$\sigma^\prime$ were not considered for
the subsequent flux determination and error calculation. The counts
were then transformed into absolute fluxes taking into account the
calibration observations made on standard stars which were used to
derive the zero point of our photometric system and the atmospheric
extinction. The associated errors were calculated using standard error
propagation analysis, yielding typical values of $\approx$\,0.03\Mag\,
for 'well-detected' sources in the three filters considered.

In Table A2 of the Appendix we present the near infrared magnitudes
(or lower magnitude limits) derived for the Arecibo OH/IR stars
observed. No photometric data are available for IRAS\,18549+0208,
because it could not be observed under photometric conditions in any
of the observing epochs, nor for IRAS\,19029+0933 and
IRAS\,20149+3440, because these sources were found to be strongly
blended with a field star. Ten objects were barely detected in the K
band, and were therefore too faint to perform reliable photometry. For
them, lower magnitude limits in the three filters are given. Because
of various technical problems we were not able to derive the K band
magnitude of IRAS\,19422+0933, and the H band magnitude of
IRAS\,20440+0412. We noticed during the data reduction process that
several bright OH/IR stars observed with the large aperture were
measured in the non-linear regime of the detector resulting in an
underestimation of their brightness. For these objects upper magnitude
limits are given.

\section {Discussion}


We made a number of consistency checks involving positional
correlations and the analysis of colours, to validate the
near-infrared counterparts selected and to single out exceptional
objects.

\subsection{Positions of the near-infrared counterparts}

We have compared our new positions: i) with those originally listed in
the IRAS Point Source Catalogue for all the sources in the sample with
an NIR counterpart; ii) with those for a subsample of 243
objects with an MSX counterpart; and
iii) with the accurate VLA radio coordinates of 44 OH/IR stars. In
Table \ref{tab_arec:diff_coord} we list the median and the mean
separation in right ascension ($\alpha$) and declination ($\delta$)
between our Calar Alto coordinates and the IRAS, MSX and VLA ones,
together with the associated standard deviations.


\begin{table}

\caption[Separation between Calar Alto and IRAS, MSX and VLA coordinates]
{\label{tab_arec:diff_coord}Median and mean separation (and standard
deviation) between Calar Alto (CA) coordinates and IRAS, MSX and VLA
ones.}

\begin{center}
\begin{tabular}[t]{lrcccc}
\hline\hline\noalign{\smallskip}
 & N & $\Delta\alpha$ & $\Delta\alpha$ ($\sigma_{\Delta\alpha}$) & $\Delta\delta$ & $\Delta\delta$ ($\sigma_{\Delta\delta}$) \\ 
 &   &    median      &   mean                                   &   median       &   mean                                   \\
\noalign{\smallskip}\hline\noalign{\smallskip}
CA--IRAS& 363 & 3.1\arcsec & 5.6\arcsec\, (7.1\arcsec) & 2.0\arcsec & 2.3\arcsec\, (2.2\arcsec) \\
CA--MSX & 243 & 1.4\arcsec & 1.5\arcsec\, (1.4\arcsec) & 1.0\arcsec & 1.3\arcsec\, (1.4\arcsec) \\
CA--VLA &  44 & 1.0\arcsec & 1.0\arcsec\, (0.8\arcsec) & 0.6\arcsec & 1.1\arcsec\, (1.4\arcsec) \\
\noalign{\smallskip}\hline
\end{tabular} 
\end{center}
\end{table}


Our new coordinates are in excellent agreement with those obtained at
the VLA, with an average offset in both RA and Dec. of
$\approx$\,1\arcsec. The only exception is IRAS\,18554+0231 with an
offset of 7\arcsec\ (cf. Table A1). Lewis et al. (\cite{Lewis90b})
list for this source a larger astrometric uncertainty than for the
others, implying that the difference is probably due to an error in
the radio position.

\begin{figure}
\begin{center}
   \resizebox{\hsize}{!}{\includegraphics{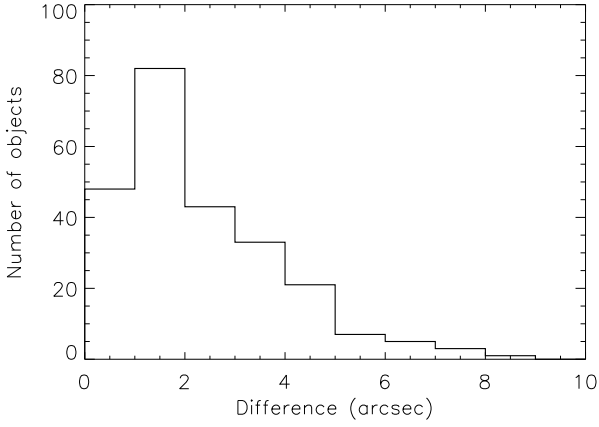}}

    \caption[Angular separation between CA and MSX
      coordinates]{\label{fig_arec:CA-MSX}\,\, Angular separation
      between Calar Alto and MSX coordinates for the 243 OH/IR stars
      in common for which an optical/near-infrared counterpart was
      found.  }

\end{center}
\end{figure}

In addition, there are 243 sources with available MSX coordinates in
our `Arecibo sample'. The angular separation between the MSX
coordinates and our new coordinates is shown in Fig.
\ref{fig_arec:CA-MSX}. The median angular separation is
$\approx$\,1.7\arcsec.
The very good positional agreement with the MSX data reinforces the
reliability of our identifications.


\begin{figure}
\begin{center}
 \resizebox{\hsize}{!}{\includegraphics{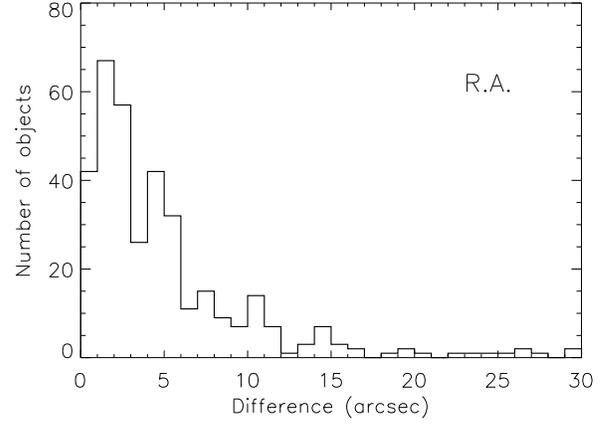}}
 \resizebox{\hsize}{!}{\includegraphics{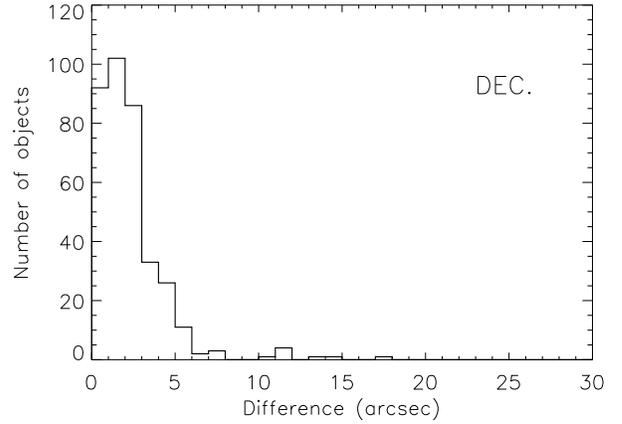}}

    \caption[$\mid$$\Delta\alpha$$\mid$ and $\mid$$\Delta\delta$$\mid$
    between CA and IRAS coordinates]{\label{fig_arec:CA-IRAS}\,\,
    Absolute coordinate differences in right ascension (upper panel)
    and declination (lower panel) between Calar Alto and IRAS
    coordinates for the 363 OH/IR stars for which an
    optical/near-infrared counterpart was found. Not included are six
    sources with
    $30\arcsec$\,$<$\,$\mid$$\Delta\alpha$$\mid$\,$<$\,$61\arcsec$.}

\end{center}
\end{figure}

As expected, much larger deviations are found when we compare our new
coordinates with the original IRAS coordinates. Misidentifications are
in this case more difficult to determine, especially if neither MSX
nor VLA coordinates are available to support our choice of a given
counterpart. Fig. \ref{fig_arec:CA-IRAS} shows the histograms of
these deviations. While in declination they are within 6\arcsec\ with
just a few exceptions, in right ascension we find a considerable
number of objects ($\approx$\,20\%) showing deviations of the order of
10\arcsec\ and even larger.
The larger errors in RA is a well known feature of the IRAS Point
Source Catalogue (Herman et al., \cite{Herman86}; Bowers \& Knapp,
\cite{Bowers89}; Lewis et al., \cite{Lewis90b}). The problem is
related to the differential accuracy of the IRAS survey in the
cross-scan and in-scan directions (Beichman et al. \cite{Beichman88}).


As part of our quality checks we verified that the differences found
between Calar Alto and IRAS coordinates were similar to those found
between MSX/VLA and IRAS coordinates (for the objects with MSX/VLA
data available).  Large but consistent deviations can be attributed to
the large errors associated to the IRAS coordinates. For the remaining
Arecibo sources (those with no MSX or VLA counterparts) it is more
difficult to verify that the coordinate deviations are not due to a
misidentification. In Table A1 there are 14
sources with extremely large differences in RA ($\ge$\,20\arcsec). 9
of them have MSX coordinates showing very similar deviations with
respect to the original IRAS coordinates. The remaining five sources
(IRAS\,18033+2229, IRAS\,19346+0913, IRAS\,20127+2430,
IRAS\,20194+1707 and IRAS\,21305+2118) are bright IRAS as well as
near-infrared sources. They also have blue IRAS colours and are
located at galactic latitudes $\mid$b$\mid$\,$>$\,5$^{\circ}$, where
confusion is unlikely. We conclude therefore that the identification
of these counterparts is correct and that deviations with respect to
the IRAS coordinates up to $\approx$\,60\arcsec\ are possible,
although not frequent, and are due to the large errors in the IRAS
astrometry.

\subsection{Near-infrared magnitude distribution of the `Arecibo sample'}


\begin{figure*}
\begin{center}
	\includegraphics[width=13cm]{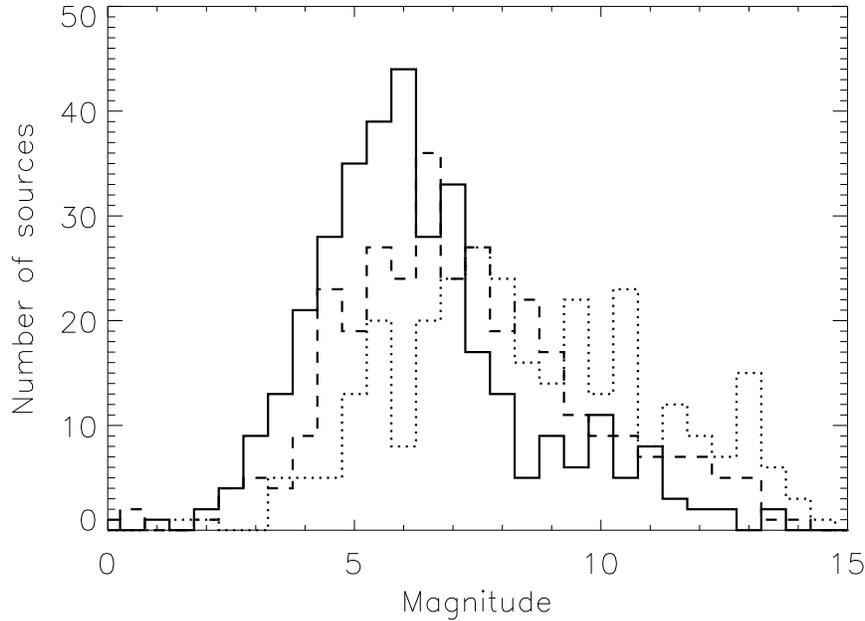} 
	\caption[Near-IR magnitude distribution of the Arecibo
	sources]{\label{fig_arec:NIR_histogram}\,\ J band ({\it dotted
	line}), H band ({\it dashed line}), and K band ({\it solid
	line}) magnitude distribution of the sources included in the
	`Arecibo sample'.}
\end{center}
\end{figure*}

Fig. \ref{fig_arec:NIR_histogram} shows the J, H, and K magnitude
distribution of the 351 Arecibo sources with a near-infrared
counterpart for which a photometric measurement was done at least
in the K band. Most of the sources observed have bright near-infrared
counterparts with the magnitude distribution peaking around 6.0\Mag\,
in K, 6.5\Mag\, in H, and 7.5\Mag\, in J. The distributions are
however not strongly peaked and have a long tail extending to the
sensitivity limit of our observations. We estimate that the
distributions in Fig. \ref{fig_arec:NIR_histogram} become incomplete in
all filters at $\approx$\,12\Mag. 18\%
of the observed sources were not detected in the J-band, and 10\%
were not detected in the H band. As already mentioned in Sect.
\ref{arec_sect_atlas}, for 18 sources (5\% of the sample) the 
counterpart was too weak to derive a reliable photometric measurement
in the K band.

The observed distribution can in principle be attributed: i) to the
different intrinsic brightnesses of the sources in the sample; ii) to
the different apparent luminosities expected from sources located in a
wide range of distances; and iii) to the different optical thickness
of their Circumstellar Envelope (CSE). From the photometric data alone
it is unfortunately not possible to disentangle the relative
contribution of these three effects for individual sources. However,
considered as a whole, we can derive some statistical conclusions from
Fig. \ref{fig_arec:NIR_histogram}. According to previous studies
reported in the literature (cf. Habing \cite{Habing96}), and the
results obtained by us (Jim\'{e}nez-Esteban et
al. \cite{Jimenez-Esteban04}) the distribution of intrinsic
luminosities expected in our sample of OH/IR stars covers the range
between $\approx$\,2\,500\,\Lsun\, and
$\approx$\,26\,000\,L$_{\odot}$. This would translate into a
dispersion in observed brightness of the order of $\approx$\,2.5\Mag.
Moreover, in Jim\'{e}nez-Esteban et al. (\cite{Jimenez-Esteban04}) we
found that the distribution of luminosities is actually strongly
peaked around 3\,500\,--\,4\,000\,\Lsun. This implies that the
dispersion in the observed brightness due to the scatter of intrinsic
luminosities must be even smaller when statistically considered. On
the other hand, in Jim\'{e}nez-Esteban et
al. (\cite{Jimenez-Esteban04}) we found that most of our sources
are estimated to be randomly distributed at distances ranging from 1
to 5\,kpc. This would mean that their apparent luminosities would
cover a brightness range equivalent to $\approx$\,3.5\Mag. The
combination of these two effects is not enough to explain the wide
range of brightness shown in Fig.
\ref{fig_arec:NIR_histogram} (more than 8\Mag\, in all filters). 
Thus, the different optical thickness of the CSE of individual OH/IR
stars must contribute significantly to the observed scatter.

\subsection{\label{arec_secJ-HvsH-K}The J--H\,vs.\,H--K colour-colour diagram} 


\begin{figure*}
   \resizebox{\hsize}{!}{\includegraphics{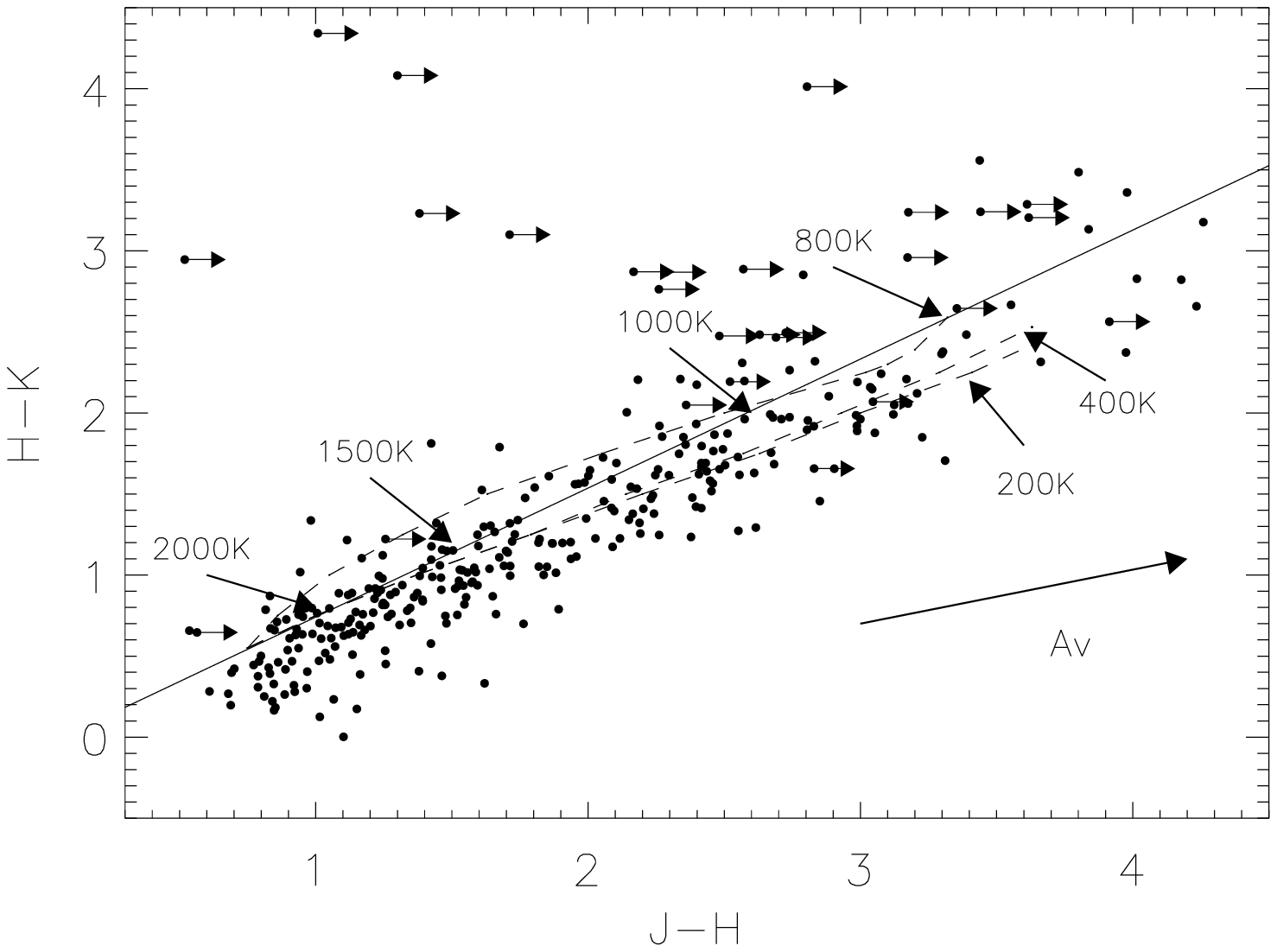}}

   \caption[J--H\,vs.\,H--K colour-colour diagram for the Arecibo
	sources]{\label{fig_arec:NIR_colour}\,\, Near-infrared
	J--H\,vs.\,H--K colour-colour diagram of all Arecibo sources
	with photometry at least in the H and K bands. J--H lower
	limits are indicated by arrows.  The solid line shows the
	location of black-bodies of different temperatures and the
	dashed lines show the location of the combination of a
	2\,500\,K black-body with dust shells of temperatures 200, 400
	and 800\,K following Whitelock (\cite{Whitelock85}). In
	addition we also indicate the reddening vector corresponding
	to A$_V$\,=\,10\Mag.}

\end{figure*}


In Fig. \ref{fig_arec:NIR_colour} we have plotted the near-infrared
colour-colour diagram J$-$H\,vs. H$-$K of all sources with a
detection at least in the H and K bands. For objects not detected in
J, lower limits for the J$-$H colours were calculated and these are
indicated with an arrow in this diagram.

The J$-$H and H$-$K colours show an almost linear correlation
extending from (J$-$H\,=\,0.5; H$-$K\,=\,0.3) up to (J$-$H\,=\,4.2;
H$-$K\,=\,3.2). The scatter in the range 0.5\,$<$\,J$-$H $<$\,3.0 is
0.26\Mag, which is a factor 5\,--\,10 larger than expected from the
errors of our photometry. Beyond J$-$H\,=\,3.0 the scatter increases
to $\approx$\,0.40\Mag.

The colour-colour distribution is similar to that obtained by Lewis et
al. (\cite{Lewis04}) from 2MASS data for one third of the
`Arecibo sample'. At its blue end it encompasses also the region where
Mira variables are found (Whitelock et
al. \cite{Whitelock94}). Unfortunately, the limited sensitivity of our
observations especially in the J band prevents the full exploration of
the red end of the distribution. If we extrapolate the correlation found,
the reddest objects in our sample with H--K\,$>$\,4\Mag\, would have
associated colours J--H\,$\approx$\,6, well beyond the limit of
J$-$H\,=\,4.2 shown in Fig. \ref{fig_arec:NIR_colour}.

In general, the position of the sources in Fig.
\ref{fig_arec:NIR_colour} can be explained as the result of the
combined emission of a cool star (T\,$\approx$\,2\,500\,K) and a much
cooler dust shell (T\,$<$\,800\,K). The main effect of the shell in
the near-infrared colours would be to increase the circumstellar
reddening, which, together with differential interstellar extinction
effects from source to source, could explain the width of the
distribution observed.

But the correlation shown in the near-infrared colour-colour diagram
J$-$H\,vs.\,H$-$K essentially represents a sequence of increasing
optical thickness of the CSEs, where Mira variables with still
optically thin CSEs are placed in the bluer part, and OH/IR stars with
thicker CSEs are located in the redder part of the diagram. We expect
therefore that the fraction of objects with optical counterparts will
decrease with increasing near-infrared colours.


\begin{figure}
	\includegraphics[width=8.5cm]{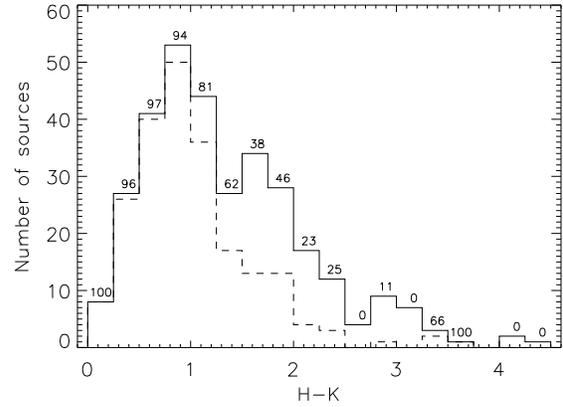}

	\caption[H--K distribution of the Arecibo sources with and
	without optical
	counterpart]{\label{fig_arec:Opt_detect_rate_H-K}\,\, The
	total number of sources in the sample ({\it solid line}) is
	compared with the number of sources with an optical
	counterpart on the DSS2 images ({\it broken line}) as a
	function of the H--K colour. The numbers on top of each bin
	indicate the percentage of optical detections.}

\end{figure}


In Fig. \ref{fig_arec:Opt_detect_rate_H-K} we have plotted the
fraction of optically identified counterparts on the DSS2 images
versus the near-infrared colour H$-$K. As expected, the percentage of
detections decreases with H--K colour. For H--K\,$\la$\,1 almost all
Arecibo sources have optical counterparts, while at H--K\,$\ga$\,1.5
the fraction decreases rapidly to $<$\,50\%. The two very blue
(H--K\,$<$\,0.7) Arecibo sources without optical counterpart are
IRAS\,18409+0431 (J\,=\,6.8\Mag) and IRAS\,19178+1206 (H\,=\,14.2\Mag)
(cf. Table A1). The non-detection of IRAS\,18409+0431 might be caused
by variability or extremely high extinction, while IRAS\,19178+1206
will be later identified in this chapter as a non-variable OH/IR star
already in the early post-AGB stage. Beyond H--K\,=\,2 the fraction of
optical detections decreases further but does not go down completely
to zero. The very red stars (H--K\,$>$\,2.5) with optical counterparts
are probably highly variable OH/IR stars optically detected during a
luminosity maximum. Another possibility is that the proposed optical
counterparts are misidentified and they are actually faint field stars
unrelated with the IRAS sources.

%

\begin{figure*}
\begin{center}
   \resizebox{\hsize}{!}{\includegraphics{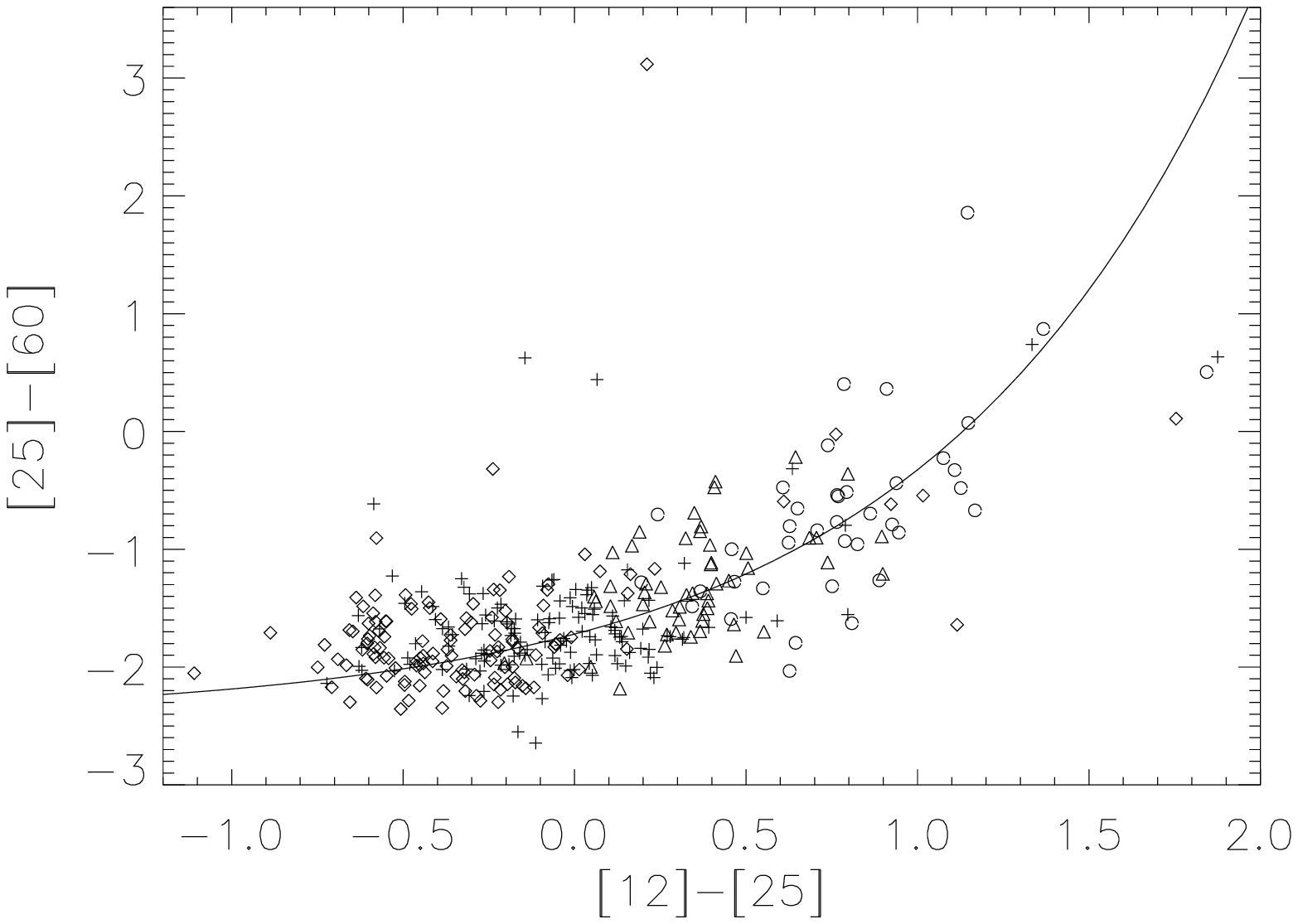}}
     \caption[IRAS two-colour diagram for the Arecibo sources as a
	  function of
	  H$-$K]{\label{fig_arec:12_25vs25_60_binsH_K}\,\, IRAS
	  two-colour diagram with the Arecibo sources distinguished
	  according to the H$-$K colours of their near-infrared
	  counterparts. Diamonds
	  ($\diamondsuit$) are sources with 
	  H$-$K\,$\le$\,1.0, crosses (+) are those with 1.0\,$<$\,H$-$K\,$\le$\,2.0,
	  and triangles ($\triangle$) those with 2.0\,$<$\,H$-$K.  
	  Circles ($\circ$) correspond to sources with H brightness
	  below the detection limit. The solid line is the `O-rich AGB
	  sequence'.}
\end{center}
\end{figure*}

Fig. \ref{fig_arec:12_25vs25_60_binsH_K} shows the position of the
Arecibo sources in the IRAS colour-colour diagram as a function of
their near-infrared colours. Blue near-infrared sources with
H$-$K\,$\le$\,1.0 are located preferentially in the bluer part of the
diagram ([12]$-$[25] $<$ $-$0.1), while those with a red near-infrared
colour (H$-$K\,$>$\,2.0), most of them heavily obscured sources
without an optical counterpart as we have just shown, occupy the red
part of the diagram ([12]$-$[25]\,$>$\,0.1). Those with intermediate
near-infrared colours occupy an intermediate range of [12]$-$[25]
colours, as expected. There are however large overlaps between these
groups.

The smooth correlation found between near- and mid-infrared colours
corroborates the generally accepted interpretation that both the H$-$K
and the [12]$-$[25] colour sequences are good indicators of the
optical thickness of the CSE. The colours cover the whole optical
thickness range, from the optically thin shells of Mira variables to
the thick shells of OH/IR stars. As the transparency of the shell
decreases, stars are expected to become obscured, initially in the
optical range and subsequently in the near-infrared at increasing
wavelengths. The very red objects which could not be detected in the H
band represent the most extreme AGB population of obscured stars. Most
of them are actually located among the reddest sources also in the
IRAS two colour-colour diagram, as we can see in Fig.
\ref{fig_arec:12_25vs25_60_binsH_K}.

The detection of a few blue near-infrared (H--K\,$<$\,1.0)
counterparts with very red [12]$-$[25] ($>$\,0.55) colours could be an
indicator of the end of the high mass loss regime in these stars,
which should then be classified as post-AGB stars. Stars in the
post-AGB phase are expected to reappear again as bright sources first
in the near-infrared and later in the optical range, while the remnant
of the circumstellar shell dissipates. Alternatively, they could also
be field stars, with the true counterparts being heavily obscured
beyond our detection limits. We will analyse this small group of stars
later more in detail.

\subsection{The K$-$[12]\,vs.\,[12]$-$[25] colour-colour diagram}

\begin{figure*}
\begin{center}	
   \includegraphics[width=13cm]{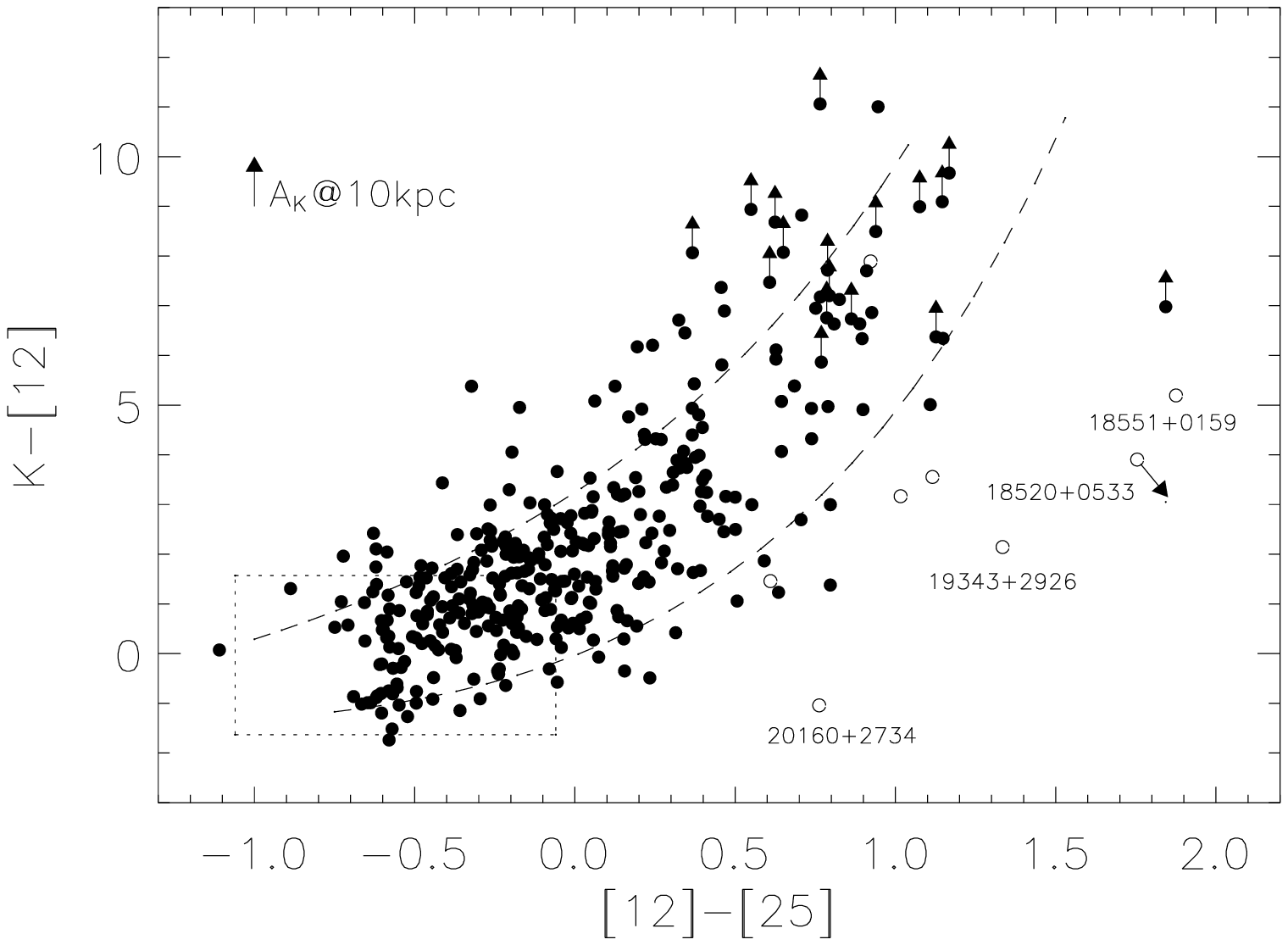}
    \caption[Near/mid-IR colour-colour diagram for the `Arecibo
    sample']{\label{fig_arec:N-MIR_colour}\,\, 
      K$-$[12]\,vs.\,[12]$-$[25] colour-colour diagram, where the
      K$-$[12] colour is defined as 
      K$-$[12]\,=\,$-$2.5\,log$\frac{F_{\nu}(2.2)}{F_{\nu}(12)}$, and
      the [12]--[25] colour is defined as in Fig.
      \ref{fig_arec:IRAS_colour}. The dashed lines indicate the
      dispersion expected from the intrinsic variability of these
      sources. Candidate post-AGB stars are marked with open
      symbols. The dotted box delimits the region occupied by the type
      of Mira variables studied by Whitelock et
      al. (\cite{Whitelock94}). For details, see text.}

\end{center}
\end{figure*}

Another way to analyse the near- and mid-infrared properties of the
sources in our sample is to study their distribution in the
K$-$[12]\,vs.\,[12]$-$[25] colour-colour diagram (Fig.
\ref{fig_arec:N-MIR_colour}). All stars with a reliable K band 
photometry and good quality IRAS fluxes at 12 and 25\,\mic\, are
plotted. If only upper limits in flux (lower limits in magnitudes) are
available these are indicated with arrows.
In stars with optically thin shells in the near-infrared the K$-$[12]
colour provides information on the relative contribution to the
overall spectral energy distribution of the near-infrared emission,
dominated by the central star and by the hot dust surrounding it, and
of the mid-infrared emission, mainly coming from the cool dust in the
circumstellar shell. The correlation of the K$-$[12] with the IRAS
[12]$-$[25] colour can then be interpreted as an additional indicator
of the optical thickness of the CSE of a given source. To calculate
the K$-$[12] colour we adopted a zero-magnitude flux in the K band of
665\,Jy (Koornneef
\cite{Koornneef83}).

Although a clear correlation exists between the K$-$[12] and the
[12]$-$[25] colours, as is shown in Fig. \ref{fig_arec:N-MIR_colour},
there is a very large scatter on both axes. Only a small part of this
scattering can be attributed to the intrinsic errors associated to the
photometric measurements, mainly in the IRAS data, as the errors
associated to our K band photometry are in most cases
negligible. Typical errors are in the 10\,--\,15\% range at 12 and
25\,$\mu$m for the sources in our sample which translates into a
dispersion of just 0.1\,--\,0.2\Mag\, in the [12]$-$[25] colour.
The influence of interstellar extinction is also expected to be low,
as is indicated by the small size of the reddening vector shown in
Fig. \ref{fig_arec:N-MIR_colour}, which corresponds to the reddening
effect expected for a star located at a distance of 10\,kpc in the
K$-$[12] colour (the effect on the [12]$-$[25] colour would be
negligible). The largest contribution to the colour dispersion
observed is expected to come from the strong variability of the
sources in our sample. In the case of the K$-$[12] colour, the fact
that the near- and mid-infrared observations under analysis were
non-contemporaneous adds a substantial scatter. From data in the
literature we know that optically bright Miras show typical K band
amplitudes of 0.5\,--\,1.0\Mag\, while the most extreme OH/IR stars
can reach amplitudes in the K band of up to 4\Mag\, (Engels et
al. \cite{Engels83}; Le Bertre \cite{LeBertre93}; Olivier et
al. \cite{Olivier01}). In addition, the bluest Miras with OH maser
emission show a variability at 12\,$\mu$m of $\approx$\,0.6\Mag\,
(Smith \cite{Smith03}) while this variability can reach up to
1.8\Mag\, in the most extreme OH/IR stars. This latter value has been
estimated by comparing the measured flux in 12\,\mic\, both by IRAS
and MSX satellites for those Arecibo sources for which these
photometric data exist. If we include all these uncertainties in the
K$-$[12]\,vs.\,[12]$-$[25] colour-colour diagram we obtain a wide area
delimited by the dashed lines plotted in Fig.
\ref{fig_arec:N-MIR_colour}, where the majority of the stars are
located. This would imply that in most cases the intrinsic variability
of the sources could explain the scatter observed.

The K$-$[12] colour distribution has also been studied by
Whitelock et al. (\cite{Whitelock94}) for Mira variables and by
Sevenster (\cite{Sevenster02}) for OH-selected AGB and post-AGB
stars. 
The Mira variables in Whitelock et al. (\cite{Whitelock94}) cover the
blue end of the J--H\,vs.\,H--K two-colour diagram (Fig.
\ref{fig_arec:NIR_colour}) with the majority of them concentrated
around values of J--H\,$\approx$\,1.0 and H--K\,$\approx$\,0.5. In
contrast, the OH-selected AGB and post-AGB stars studied by Sevenster
(\cite{Sevenster02}) show a wider distribution in this diagram, very
similar to what we observe in our `Arecibo sample'.


In the K$-$[12]\,vs.\,[12]$-$[25] two-colour diagram the blue
oxygen-rich Miras studied by Whitelock et al. (\cite{Whitelock94}) are
confined to the region indicated in Fig.
\ref{fig_arec:N-MIR_colour}. In the Arecibo sample, K$-$[12] values are
in many cases considerably redder, in agreement with the values found
in the OH-selected sample of AGB and post-AGB stars used by Sevenster
(\cite{Sevenster02}). The scarcity of sources with very red K$-$[12]
colours in Whitelock et al. (\cite{Whitelock94}) is therefore largely
a selection effect.


For Mira variables the K$-$[12] colour is interpreted by Whitelock et
al. (\cite{Whitelock94}) as a measure of the mass-loss rate arguing
that the K and 12\,\mic\, fluxes originate respectively from the star
and the shell. This argument is, however, alleviated with the
increasing optical thickness of the shell and the corresponding
diminishing contribution from the star. The observed range of K$-$[12]
colours for a given [12]$-$[25] value is therefore a result of several
contributions and a more thorough study is needed to properly explain
the observations. This will be possible as soon as the K-band
variability can be removed from the diagram as the result of our
on-going monitoring program.

\subsection{Early post-AGB candidates}

Fig. \ref{fig_arec:N-MIR_colour} also shows at least four clear
outliers displaying a near-infrared counterpart much brighter than
expected from their red [12]$-$[25] colours. They are labelled with
their IRAS name in the K$-$[12]\,vs.\,[12]$-$[25] diagram. Namely,
viz. IRAS\,18520+0533, IRAS\,18551+0159, IRAS\,19343+2926 and
IRAS\,20160+2734. The most natural interpretation for their peculiar
position in this diagram is that they are already in the post-AGB
stage. As we have already mentioned, once the large mass loss at the
end of AGB evolution ceases, the optical depth of the circumstellar
envelope decreases and the central star is expected to reappear again
at shorter wavelengths while the mid-infrared colours redden further
for some time (Bedijn \cite{Bedijn87}; Vassiliadis \& Wood
\cite{Vassiliadis93}). The cessation of the high mass loss is
connected to the end of the large amplitude pulsation of regular OH/IR
stars. Thus, non-variability is expected to be an additional post-AGB
stage indicator.

In addition, we have also indicated with open symbols in
Fig. \ref{fig_arec:N-MIR_colour} the position of those sources with
very red [12]$-$[25] colours and unusually blue near-infrared
counterparts that were also tentatively classified as candidate
post-AGB stars or proto-planetary nebulae in
Sect. \ref{arec_secJ-HvsH-K}. As we can see, two of these sources,
IRAS\,18520+0533 and IRAS\,20160+2734, are among the outliers that we
have just identified as candidate post-AGB stars, which supports our
classification.
 
Both samples of candidate post-AGB stars are listed together in Table
\ref{tab_arec:pec_objects}, where we give their photometric colours H$-$K,
K$-$[12] and [12]$-$[25], the IRAS variability index and the proposed 
identification.



\begin{table*}

\caption[Early post-AGB candidates]{\label{tab_arec:pec_objects}Very
red sources with unusually bright and/or blue near-infrared
counterparts identified as candidate post-AGB stars or proto-planetary
nebulae.}

\begin{center}
\begin{tabular}[t]{lcrcrl}
\hline\hline\noalign{\smallskip}
IRAS      & H$-$K& K--[12] & [12]--[25]  & VAR & Comments \\
\hline\noalign{\smallskip}\noalign{\smallskip}
18095+2704&  0.26&3.55 & 1.11 &  15 &  Post-AGB star \\
18520+0533&  0.93&3.90 & 1.76 &   3 &  Peculiar OH/IR star\\
18551+0159&  1.66&5.20 & 1.87 &  25 &  Infrared Planetary nebula\\
19065+0832&  0.70&3.17 & 1.02 &  98 &  Non-variable OH/IR star?\\
19178+1206&  0.65&7.89 & 0.92 &   1 &  Non-variable OH/IR star \\
19200+1035&  0.82&1.46 & 0.61 &   0 &  Planetary nebula\\
19343+2926&  1.79&2.15 & 1.33 &   0 &  Bipolar proto-planetary nebula\\
20160+2734&  0.29&$-$1.04&0.76&   0 &  Semi-regular variable \\
\noalign{\smallskip}\hline
\end{tabular} 
\end{center}
\end{table*}


As we can see, several of these stars have already been identified as
post-AGB stars, proto-planetary nebulae or even as planetary nebulae
in the literature.


This is the case for IRAS\,18095+2704 a well-known
high-galactic-latitude post-AGB star, with spectral type F3Ib (Hrivnak
et al. \cite{Hrivnak88}), for which a detailed spectroscopic analysis
including determination of chemical abundances can be found in
Klochkova (\cite{Klochkova95}).

%

The NIR photometry of IRAS\,18520+0533 shows no variability between
our two observing epochs and it has a low IRAS variability index.
Eder et al. (\cite{Eder88}) found that this source has a peculiar
1612\,MHz spectrum. Its two strongest peaks show the standard profile
observed in OH/IR stars, but exterior to these there is a pair of
small flat peaks which may be due to a second shell of circumstellar
material expanding at a larger velocity. This points to the peculiar
nature of this OH/IR star, but we note that its location in the
K$-$[12]\,vs.\,[12]$-$[25] diagram depends strongly on the IRAS
12\,\mic\, flux, for which there is only an upper limit in the IRAS
PSC.

IRAS\,18551+0159 was first identified as an infrared planetary nebula
by Kistiakowsky \& Helfand (\cite{Kistiakowsky95})
on the basis of its strong radio continuum emission at 20\,cm and the
large
F([S\,III]\,$\lambda$\,9\,532\AA)\,/\,F(Paschen\,9\,$\lambda$\,9\,225\AA)
ratio, which means that classification as an ultracompact H\,II region
must be discarded. The source is extremely red and has no optical
counterpart on the DSS2 image, which is probably a signature of its
recent departure from the AGB.

IRAS\,19065+0832 (OH\,42.6+0.0, GLMP\,876) is the only source in Table
\ref{tab_arec:pec_objects} showing a large IRAS variability index. It was
classified in the GLMP catalogue as a variable OH/IR star (Garc\'\i
a-Lario \cite{Garcia-Lario92}) on the basis of its large IRAS
variability index, together with the detection of a strong silicate
absorption feature in its IRAS Low Resolution Spectrum (it is a `class
39' source). However, this is not consistent with the blue colours
observed in the near-infrared nor with the non-variability which is
suggested by the very similar photometry in our two epochs (only
0.07\Mag\, difference in the K-band). Our suggestion is that, if the
near-infrared source was correctly identified, the star must be
already in the post-AGB stage, but we cannot discard the possibility
of a misidentification. This is why our classification in Table
\ref{tab_arec:pec_objects} is only tentative.


We identify IRAS\,19178+1206 (GLMP\,899) as a non-variable OH/IR star
in a very early post-AGB stage. The source is invisible in the optical
while it has a relatively blue near-infrared counterpart. The
non-variability and the blue near-infrared colours are confirmed with
our two different epochs. However, the near-infrared counterpart of
this source might in fact be a field star, and the true IRAS source
invisible even in the K band.

IRAS\,19200+1035 (K\,3$-$33) is included as a compact planetary nebula
in the Strasbourg-ESO Catalogue of Galactic Planetary Nebulae (Acker
et al. \cite{Acker92}) although with a wrong finding chart. It is
clearly identified as a bright source in our near-infrared images
while it is very faint in the optical. Again the non-variability and
the blue near-infrared colours are confirmed with our two different
observing epochs.

IRAS\,19343+2926 (M\,1$-$92), also known as Minkowski's Footprint,
is a very well known young, bipolar proto-planetary nebula (Bujarrabal
et al. \cite{Bujarrabal98}). The photometry for this source shows a
very low variation between our two observing epochs ($<$\,0.04\Mag\, in K).

IRAS\,20160+2734 (AU\,Vul) is classified as a semi-regular pulsating
star in the SIMBAD database.  It was first considered a post-AGB star
by Jim\'enez-Esteban et al. (\cite{Jimenez-Esteban01}) during a first
analysis of our present sample. Lewis et al. (\cite{Lewis04})
supported this classification on the basis of the erratic detection of
its different masers, being detected in the 1612\,MHz OH maser at the
sensitivity limit of the telescope by Eder et al. (\cite{Eder88}),
non-detected later in a water maser survey by Engels \& Lewis
(\cite{Engels96}) or as a mainline OH maser by Lewis (\cite{Lewis97}),
and recently again detected at 1667\,MHz but not at 1665\,MHz by Lewis
et al. (\cite{Lewis04}).

\section {Conclusions}

We have presented an atlas of optical/near-infrared finding charts and
near-infrared photometric observations for 371 objects taken from the
`Arecibo sample of OH/IR stars'. Except for 8 sources, we successfully
identified their near-infrared counterparts and determined new
positions for each source in the sample with an accuracy of
$\approx$\,1\arcsec.  The correctness of the identifications was
carefully assessed using improved positional information (MSX, VLA),
analysing the near- and mid-infrared colour measurements, and in some
cases searching for variability. The few sources for which no
near-infrared counterpart was found were identified either as heavily
obscured variable OH/IR stars at the very end of the AGB or as
non-variable OH/IR stars in a very early post-AGB stage.


The wide dispersion in J, H and K magnitudes is attributed not only to
the different apparent brightnesses expected for sources located in a
wide range of distances and to their different intrinsic luminosities
but also significantly to differences in the optical thickness of the
circumstellar shells.

The distribution of the Arecibo sources in the near-infrared
J$-$H\,vs.\,H$-$K colour-colour diagram is interpreted as a sequence
of increasing optical thickness of their CSEs, where Mira-like
variables with still optically thin shells are lie in the blue part of
this diagram while the more extreme OH/IR stars with thicker shells
are located in the redder part of the diagram. Their near-infrared
colours can be reproduced with a combination of the emission coming
from a cool central star (T $\approx$\,2\,500\,K) and from a much
cooler dust shell (T\,$<$\,800\,K). The dispersion of colours observed
along the sequence can be explained as differential circumstellar
and/or interstellar reddening.

This interpretation was confirmed by a determination of the detection
rate of optical counterparts as a function of the H$-$K colour. We
found that most of the Arecibo sources with H$-$K\,$<$\,1.0 are
detected in the optical range while those with H$-$K\,$>$\,2.0 are
heavily obscured sources with no optical counterpart in the DSS2
images. These sources are also located in different regions of the
IRAS [12]$-$[25]\,vs.\,[25]$-$[60] colour-colour diagram, suggesting
that both near- and mid-infrared colours are good indicators of the
optical thickness of the shell.

The connection between the near- and the mid-infrared data was
analyzed with the help of the K$-$[12]\,vs.\,[12]$-$[25] colour-colour
diagram. We found a clear correlation between these two colours which
can also be interpreted as an indication of the increasing
contribution to the overall spectral energy distribution of the
mid-infrared component (emission from the cool dust in the
circumstellar shell) with respect to the near-infrared component
(emission from the central star plus the hot dust surrounding it) as
these stars evolve along the `O-rich AGB sequence'. However, the
scatter of this correlation is found to be very large, partly because
of the strong variability of these sources in the near- and
mid-infrared and the non-contemporaneity of the observations under
comparison. The expected variability goes from 0.5 to more than
4\Mag\, in the K band and from 0.5 to 1.8\Mag\, at 12\,$\mu$m, both
increasing toward redder [12]$-$[25] colours. Once these uncertainties
are considered, the distribution observed in the
K$-$[12]\,vs.\,[12]$-$[25] diagram can be explained as a consequence
of the different optical thicknesses of the CSEs,
with the exception of a few outliers which were found to be much
brighter than expected in the near-infrared according to their
[12]$-$[25] colour.

The few sources showing very red [12]$-$[25] colours in combination
with unusually blue near-infrared counterparts as well as the outliers
in the K$-$[12]\,vs.\,[12]$-$[25] diagram have been identified as
candidate post-AGB stars (some of them are well known proto-planetary
nebulae or planetary nebulae).

The proper identification of the IRAS counterparts forms the basis for
a successful completion of our long-term infrared and optical
monitoring program begun in 1999, which is aimed at determining the
variability properties of all the sources in the `Arecibo sample of
OH/IR stars' and combine the information collected in the
near-infrared with data available at other wavelengths in order to
study more in detail the photometric behaviour of these stars in the
context of stellar evolution.

\begin{acknowledgements}

Based on observations collected at the German-Spanish Astronomical
Centre, Calar Alto, operated jointly by the Max-Planck-Institut f\"ur
Astronomie and Instituto de Astrof\'\i sica de Andaluc\'\i a (CSIC).
This work has been supported by the Deutsche Forschungsgemeinschaft
through travel grants (En~176/24-1 and 25-1) and by the Spanish
Ministerio de Ciencia y Tecnolog\'\i a through grant
AYA2003$-$09499. Part of the data reduction was supported by a grant
of the Deutscher Akademischer Auslandsdienst (DAAD) to FJE. This
research has made use of the SIMBAD database, operated at CDS,
Strasbourg, France. We acknowledge also the use of the Digitized Sky
Survey, based on photographic data obtained using the UK Schmidt
Telescope. The UK Schmidt Telescope was operated by the Royal
Observatory Edinburgh, with funding from the UK Science and
Engineering Research Council, until 1988 June, and thereafter by the
Anglo-Australian Observatory. Original plate material is copyright (c)
of the Royal Observatory Edinburgh and the Anglo-Australian
Observatory. The plates were processed into the present compressed
digital form with their permission. The Digitized Sky Survey was
produced at the Space Telescope Science Institute under US Government
grant NAG W-2166.

\end{acknowledgements}


\newpage

\appendix

\end{document}